# Nearby sources in the transition region between Galactic and Extragalactic cosmic rays

**L G Sveshnikova[1], E E Korosteleva[1], L A Kuzmichev[1], V A Prosin[1], V S Ptuskin[2], O N Strelnikova[1]**

[1] Skobeltsyn Institute of Nuclear Physic of MSU, Moscow, Russia
[2] Pushkov Institute of Terrestrial Magnetism, Ionosphere and Radio Wave Propagation of the RAN, Troitsk, Moscow Region, Russia

E-mail: sws@dec1.sinp.msu.ru

**Abstract.** In this paper, a probable interpretation of a remarkable fine structure of all particle spectra between the knee and the ankle, as well as a high content of heavy nuclei around $10^{17}$ eV measured recently in Tunka-133 and KASCADE Grande experiments, is presented in the model where production of cosmic rays in Type Ia SNRs provides observed cosmic rays (CR) flux around the knee. Subtracting the contribution of these sources from all particle spectrum we obtained the residual flux of CR in the transition region between Galactic and Extragalactic CR in the range $10^{17} \div 5 \ 10^{18}$ eV. The obtained spectrum also has a pronounced knee at energy $(2 \div 3)10^{17}$ eV and slopes $\gamma_1 \sim 1.8 \pm 0.3$ below it and $\gamma_2 \sim 3.4$ above it. We analyzed the possible contribution from known Galactic sources and found that formally the best candidate to contribute significantly to the transition region is Cas A. This source posses a number of unusual properties and considered as Type IIb SNR. However we showed that the hypothesis of extragalactic origin of CR starting from $10^{17}$ eV seems more realistic.

## 1.Introduction

The remarkable fine structure of all particle spectra above the knee was measured in Tunka-133 [1,2], KASCADE Grande [3,4], and Gamma [5] experiments. It consists of a "hardening" around $2 \times 10^{16}$ eV and the knee-like structure [1,3] (or even the bump-like structure [1,5]) around $10^{17}$ eV. One of the possible interpretations of this structure is a "single source", which determines the shape of the knee, as proposed by A. Erlykin and A. Wolfendale many years ago, and appears to be confirmed by this data [6,7]. The alternative explanation [8,9] is that the knee is provided by the contribution of the group of remnants of SNIa – named "standard candles" in Cosmology. These remnants may give the same maximum energy of accelerated cosmic rays and can provide the sharp knee [8]. To investigate the validity of "single source" and "standard candles" hypotheses taking into account the stochastic nature of sources (as in [9]) we applied in our calculations a semi-statistical approach [9]. All potential CR sources (supernova remnants) were divided into 2 groups: actual nearby young sources selected from the latest gamma astronomy catalogue; and other sources that were considered as a random population with a birth rate of 1/50 yr, distributed within the thin Galactic disk. The particle

This work is supported by RFBR grants 10-02-01443-a, 11-02-91332-NNIOa, and Russian Ministry of Education and Science Contracts № 16.518.11.7051, № 14.518.11.7046

propagation was described in terms of Green's functions with a diffusion coefficient $D \sim E^{1/3}$. Time dependent emission was taken into account in accordance with [10]. To investigate the contribution of young nearby SNRs around the knee, we simultaneously analysed in [9] both the fine structure of the spectrum around the knee (see figure 1), and the anisotropy within the fixed sample of hypothetical nearby sources.

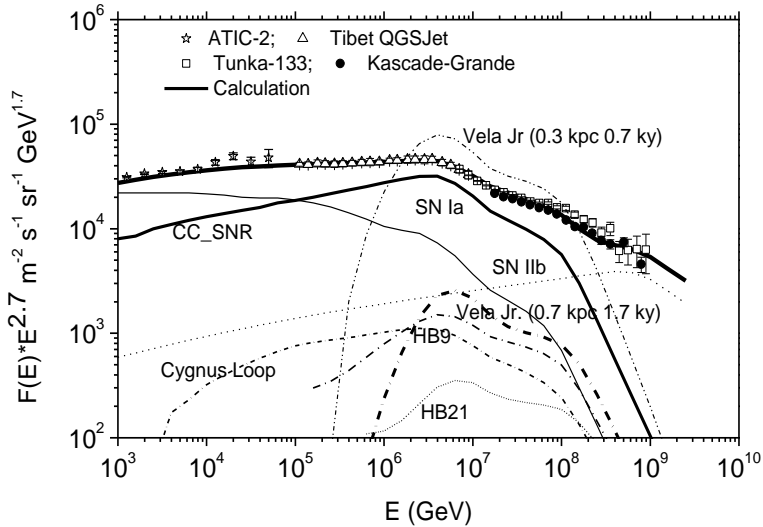

**Figure 1.** Contribution of CR from different SNR types and contribution of nearby individual sources to the all particle spectrum in the present model (see the text) together with selective experimental data ATIC-2[14], Tibet[15], Tunka 133[1], KASCADE Grande [3].

It was shown that only one source among those nearby and young enough, Vela Jr., is a good candidate to be the knee producer, but only if it is located very closely to the Earth: ~ 300 pc and T~ 700 yr (most estimations give a larger distance and age), but this suggestion contradicts small anisotropy measured around the knee [9]. Therefore, we have stated that the second hypothesis, the group of sources with more or less unified $E_{max} \sim 4 \times Z$ PeV in all remnants, looks more probable. New data concerning the spectra of heavy nuclei in the interval $10^{16}$-$10^{17}$ eV obtained in the Tunka 133 [2] and KASCADE Grande [3] experiments seem to confirm this hypothesis as will be shown below (see figure 2). The goal of this paper is to subtract the contribution of sources mainly providing the knee from all particle spectrum and to analyze the residual flux searching for unusual Galactic sources being able to provide CR in the region beyond $10^{17}$ eV [11].

## 2. The model of the sources.

In principle, the observation of high energy photons provides a direct view of the astrophysical accelerators of charged particles [12]. But the time of observation for gamma-rays and cosmic rays is shifted. This shift is ~$10^5$ yr at TeV- energies for protons emitted from sources located at 1 kpc, comparable with the life time of an SNR shell and PWN. At such a late stage of evolution the envelope disperses in ISM and a core-collapsed SN can be seen only as a pulsar or sometimes as a PWN. At energy ~ PeV a propagation time from 1 kpc decreases to ~ $10^4$ yr, in principle we can see remnants emitting PeV particles. However, we should extrapolate the evolution of the remnants back. Only at energies around $10^{18}$ eV is the diffusion time comparable with the time of light propagation, and we should directly see the process of acceleration. Taking this into account, we compiled a list of the nearest sources from the following catalogues: Green catalogue of 274SNRs; catalogue of 54 PWNe; Fermi-LAT catalogue of 46 gamma-pulsars; ATNF catalogue of 1827 pulsars; catalogue of

TeV sources of HESS, Veritas and Milagro with measured distances and the ages of sources. The total number of selected sources with known ages and distances is 73, with 24 sources inside the circle $r_i < R_{near} = 1.5$ kpc with age $t_i < T_{near} = 10^5$ yr. At a birthrate 1/50yr the expected number coincides well with the found number.

### 3. Spectrum around the knee

As the main model of cosmic ray accelerators in the Galaxy, the model developed in [8] was used. The maximum particle energy $E_{max}$ differs between four types of supernovae: $E_{max} \sim 4 \times Z$ PeV for SNIa, $E_{max} \sim 1 \times Z$ PeV for SNIb/c, $E_{max} \sim 0.1 \times Z$ PeV for SN IIp, and $E_{max} \sim (300\text{-}600) \times Z$ PeV for Type IIb SN. This model well reproduces the general shape of all particle spectrum up to more than $10^{18}$ eV, including the smooth change of the slope around the knee under the assumption of continuous distributions of sources in space and time [8].

We employ the general results of [8] in addition to introducing the discreteness of sources as explained above. Besides, more realistic case of non-fixed $E_{max}$ for cosmic rays accelerated in core collapsed supernovae was considered, as only SNRIa (~20% of all SNRs) have a small dispersion in explosion parameters and $E_{max}$ may be well fixed for them. Other SNe demonstrate a great variety in their properties, for example some core collapsed SNR (~20%) as Crab nebula has no evident shock front despite of its young age. We assumed randomly distributed $E_{max}$ for SNIIp and SNIbc in the range 1 TeV - 4000 TeV with the following reference points: 30% among all SNR have 1 TeV$<E_{max}<$10 TeV, 20% have 10 TeV$<E_{max}<$100 TeV, 20% have 100 TeV$<E_{max}<$4000 TeV. 3-5 % of all SNR have $E_{max} \sim 6 \times 10^{17}$ eV. The decreasing number density of core collapsed SNRs capable to accelerate CR up to the energy E leads to a more steep total CR source spectrum by $d\gamma \sim 0.17$ in comparison with the spectrum of individual source, $\gamma_{source}=2.2$ (it was chosen in accordance with [13]). The expected near-Earth spectrum has the slope $\gamma_{obs} = \gamma_{source} + d\gamma + d\gamma_{prop} = 2.7$ ($d\gamma_{prop} = 0.33$ is a steepening caused by propagation in the Galaxy). Calculated all particle spectrum is presented in figure 1 together with some selective data [1,3,14,15].

Here it worth noting that the fine structure at the knee region and above the knee is reproduced only if the relatively sharp cut-off in the sources' spectrum of SN Ia is assumed: the slope is changed from 2.2 to 4.2 or more. In the model, slopes of all species are similar and $E_{max} \sim Z$. The behaviour of different nuclear components around the knee can be seen in figure 2. The pink line marks the total Galactic CR with limit $Z \times 4$ PeV, other lines mark different species with chemical composition 28% of protons, 28% of He nuclei, 10% of CNO nuclei, 10% of Si nuclei, and 20% of Fe nuclei. It in the first approximation reproduces the all particle spectrum and heavy component spectra as measured by KASCADE Grande [3,4] and Tunka 133 [1,2] up to $10^{17}$ eV (heavy nuclei [*]). The change of the slope of all particle spectra $\Delta\gamma$ in this model is mainly determined by the chemical composition of the particles around the knee: $\Delta\gamma \sim \log10(\Delta Fe)/\log 26 = 0.5$ if $\Delta Fe \sim 20\%$. Needed excesses of Fe nuclei (in comparison with lower energies) in cosmic rays accelerated in SN Ia can be related to particles accelerated on the reverse shock propagating through heavy ejecta [13]. So, Tunka 133 and KASCADE Grande data confirmed long being in use idea about rigidity dependent knee.

### 4. Nearby sources providing the transition region

For the analysis of the transition region between Galactic and Extragalactic cosmic rays we subtracted the contribution of the CR accelerated in SN Ia (pink line in figure 2) from all particle spectra measured in Tunka 133 [1] and KASCADE Grande [3] experiments and obtained the residual fluxes (black points and red points correspondingly in figure 3). The residual fluxes in both cases reveal a pronounced knee with $E_{2knee} \sim 2 \times 10^{17}$ eV, a slope before the knee $\gamma_1 = 1.8 \pm 0.3$, and a slope above the knee $\sim \gamma_2 = 3.4 \pm 0.3$. The small value of $\gamma_1$ follows from a sharp cut-off of the pink line, this cutoff in their turn is necessary to explain the fine structure around the knee. The change of slope is much stronger than in the main knee: $d\gamma \sim 1.7$. If we assume in this case the rigidity dependent structure of around the $E_{2knee} \sim 2 \times 10^{17}$ eV, then we should assume a very light composition P (70%), He(25%),

+CNO(1.5%), Si-Ca (1.5%), and Fe (1%). The Fe knee in this case should be at the energy $5 \times 10^{18}$ eV, practically at the position of the "dip".

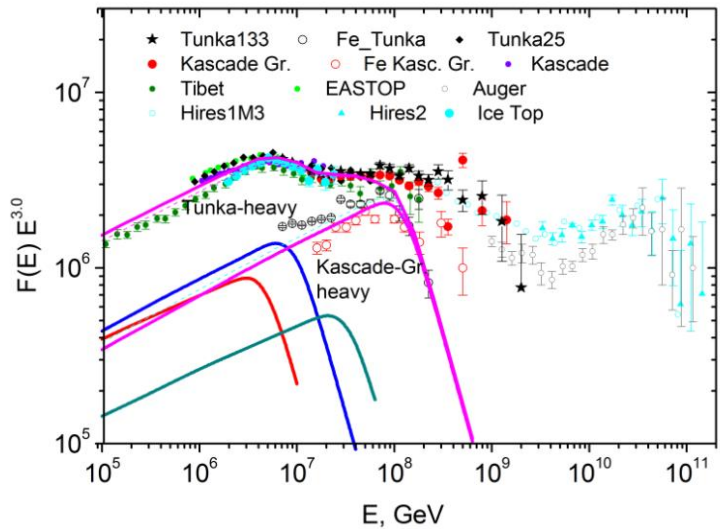

**Figure 2.** The fine structure of all particle spectrum, measured in different experiments: KASCADE Grande [3,4], Tunka 133 [1,2], Tunka_25 [16], KASCADE [18], EASTOP [17], Tibet [15] together with data, obtained in Hires 1 [19], Auger [20]. Lines denote predictions, calculated in the model described in the text, when SNIa provides the main contribution of CR around the knee: all particles (pink), protons (red), He (blue), CNO (green), Fe (light pink), z>14 (dashed blue line)

Next questions: how many sources are needed to contribute to this region? Where should the nearest one be? Can we point out a candidate? In accordance with calculation [8], only a tiny ratio (~3%) of supernova remnants with a small ejected mass, a large shock velocity, a high kinetic energy of ejecta (as IIb remnants) can provide the region of all particles' spectra around $10^{17-18}$ eV. At this energy the time of cosmic ray propagation is comparable with the time of light propagation (it takes ~10 kyr for protons to reach the Earth from the distance 3 kpc). The process of CR acceleration should be seen practically in real time. So if one assumes a birth rate of 1/50 yr for all SNRs, for these special SNRs a birth rate should be one per ~1500 yr in the whole Galaxy, and we can expect only one source with an age less than 30000 yr in the nearest region 3 kpc. Due to the rarity of these events a overwhelming contribution from the one source is expected. As an example in figure 3, the contribution of young Galactic sources, Cas A is shown for the following parameters of assumed source spectrum: $\gamma_1=2.02$, $d\gamma=2$, $E_{max}=2 \cdot 10^{17}$ eV, chemical composition: 0.7 (protons), 0.24 (helium), 0.030 (CNO nuclei), 0.028 (Si group), 0.02 (Fe group); double power for CR production in comparison with an average SNR was assumed. Cas A is the only one from the young Galactic SNRs considered as a type IIb SNR [12], similar to SN1993J. In comparison with other young SNRs, it possesses a number of unusual properties, one of which is a strong bipolarity referred as "the jet" [12], explained with a binary star scenario whereby a high mass loss is caused by a common envelope phase. In Cas A there are many iron-rich knots, indicating that the mean velocity of these knots is higher than 7000 km/sec. However, Cas A is now considered doubtful to be a candidate to the role of CR nuclei producer up to higher energies because the visible content of freshly accelerated CR in Cas A now is very small [12]. There are no other visible candidates for the Galactic source being able to accelerate CR up to $10^{17}$ eV. Other two week points of this hypothesis are the predicted heavy CR composition at energy $10^{18}$ eV and high level of anisotropy. Existing experimental data are in contradiction with these predictions.

The alternative hypothesis [11,21] that the extragalactic component occurs just at the region beyond the $10^{17}$ eV (that means $E_{max} \sim 4 \times Z$ PeV to be an upper limit of CR acceleration in Galactic sources) seems more probable. In the figure 3 we present a variant of calculation [21], where the authors studied the possible existence of the magnetic horizon: low energy protons (<$10^{17}$ eV) diffuse on extragalactic magnetic fields and cannot reach the observer within a given time due to the expansion of the Universe. Presented in figure 3, variant corresponds to the average strength of magnetic field $B_0=2$ nG, coherence length $l_c=100$ kpc, gas density $n=10^{-5}$ Mpc$^{-3}$: all of these parameters are poorly understood and poorly constrained. However, a relatively sharp cutoff of the low energy part of extragalactic proton spectrum is predicted [22]. Together with a predicted sharp steepening of the Galactic spectrum around the energy $4\times26$ PeV in our model, we can expect that some structure, "bump" or "second knee", can appear at the border between Galactic and Extragalactic CR [22]. The main signature of this hypothesis is a light composition in the region $2\times10^{17}$-$10^{18}$ eV.

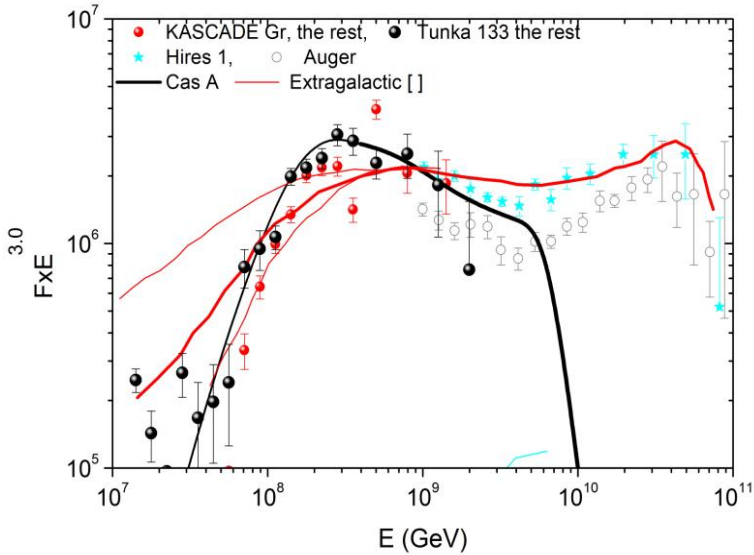

**Figure 3.** The residual flux of particles obtained after subtraction of the contribution of knee particles from the Tunka-133 data (black balls), and from KASCADE Grande data (red balls). Other points: Hires 1 [19], Auger [20]. Thick black line: expected contribution of Cas A; parameters of source spectrum: $\gamma_1=2.05$, $d\gamma=2$, $E_{max}=2\ 10^{17}$ eV. Red lines: calculated extragalactic cosmic rays from [21].

5. **Conclusions.**

We suggested a model of Galactic cosmic ray sources [9] that allowed to take into account known information about real nearby supernova remnants from latest gamma-catalogues and investigate the possible candidates for the sources giving the main contribution around the knee region, and the transition region between $10^{17}$-$5\ 10^{18}$ eV.

1) The fine structure of all particle spectra between the 1$^{st}$ knee and the point $\sim10^{17}$ eV, as well as a high content of Fe nuclei around $10^{17}$ eV, can be reproduced in this model by the class of sources SNIa with a birth rate of $\sim20\%$ from all SNR with $E_{max} \sim Z$ 4-5 PeV; chemical composition P+He$\sim$ 60%, Fe$\sim$11-13%; a sharp cutoff of source spectrum above $E_{max}$ by $d\gamma \sim2$.
2) Only one source, Vela Jr., if it is very young and close (R$\sim$0.3-0.5 kpc, T$\sim$0.7-1.7 ky) is suited to the role of the "single source", which determines the "sharpness" of the knee, but the background should also be abundant in Fe nuclei in the region of $10^{17}$ eV. This in turn implies the rigidity dependent knee around 4 PeV for background sources.
3) After subtraction of contribution of sources which produced the knee from all particle spectra measured in Tunka 133 [1] and KASCADE Grande [3] data, we obtained the residual fluxes of all particles, that also reveal a pronounced knee at E$\sim$2$\times 10^{17}$ eV, with a strong change of the spectral slope at this point, $d\gamma\sim1.7$.
4) The residual cosmic ray flux in the region $10^{17}\div5\times10^{18}$ can be interpreted as CR accelerated in one relatively distant Galactic source with $E_{max} \sim Z\times(2\div3)10^{17}$, and with a source spectrum $\gamma \sim2.0$, the best candidate is Cas A, that possesses number of unusual properties, and considered as a Type IIb SNR. However, this hypothesis predicts an absence of protons and helium nuclei and the high level of anisotropy at energy $10^{18}$ eV, that is in contradiction with existing data.
5) The alternative interpretation when the transition to extragalactic component occurs at the region just above $10^{17}$ eV (that means $E_{max}$=~4×Z PeV is an upper limit of CR acceleration in Galactic sources) seems to be more probable. Data in this case confirms the strong cutoff of low energy protons (<$10^{17}$ Ev) caused by a "magnetic horizon" effect, predicted in [21].